# Performance of a 967 nm CW diode end-pumped Er:GSGG laser at 2.79 μm

Z.H.Wu[1,2,3], D.L. Sun[1], S.Z. Wang[1], J.Q. Luo[1], X.L. Li[1], L. Huang[1], A.L. Hu[1], Y.Q. Tang[1], Q. Guo[1]*

[1]Anhui Institute of Optics and Fine Mechanics, Chinese Academy of Science, Hefei, 230031, China

[2]College of electro-optics, Nanjing University of Science and Technology, Nanjing, 210094

[3]ChinaAcademy of Engineering Physics, Mianyang, 621999

**Abstract**: We demonstrated a 967 nm diode end-pumped Er:GSGG laser operated at 2.794 μm with spectrum width 3.6 nm in the continuous wave(CW) mode. The maximum output power of 440 mW is obtained at an incident pumping power of 3.4 W, which corresponds to an optical-to-optical efficiency of 13% and slope efficiency of 13.2%. The results suggest that short cavity and efficient cooling setup for crystal are advantageous to improve laser performance.

## 1. INTRODUCTION

Laser with $Er^{3+}$ doped crystal emitting 2.7-3 μm has attracted considerable attention for its potential medicine application due to strong water absorption in this waveband [1]. Additionally, 2.7-3 μm is also a preferable pumping source for 3-13 μm optical parameter oscillation (OPO) laser [2-4]. Lasers around 2.8 μm have been reported in various materials such as Er:YAG, Er:YSGG, Er:GGG and Er:YLF. Especially, the Er:YSGG and Cr,Er:YSGG lasers pumped by flashlamp, Ti:sapphire and diode have been widely studied [1, 5-14], showing higher slope efficiency and lower threshold among these crystals. However, R.C.Stoneman et al reported that a highest slope efficiency of 36% was obtained on Er:GSGG crystal pumped by 970 nm Ti:sapphire laser [15], comparing with 31% on Er:YSGG crystal pumped by Ti:sapphire[6]. Furthermore, Er:GSGG can still maintain its optical characteristics almost unchanged after 100 Mrad gamma radiation [16], suggesting possesses excellent radiation resistant property and can be applied in harsh radiation environments, such as outer space. Thus, the 2.8 μm laser performance of Er:GSGG crystal is deserved to investigate deeply. Despite Ti:sapphire pumped 2.8 μm Er:GSGG laser has exhibited higher slope efficiency and lower threshold[6], but unfortunately, its 2.8 μm laser output power is limited owing to low power of Ti:sapphire laser when it is tuned at 970 nm. On the contrary, in recent years, diodes around 970 nm with high power (>20 W) have been well-developed and commercialized at low price. Therefore, it is promising to obtain high power laser at 2.8 μm with diode pumping. Here we report a 2.79 μm Er:GSGG laser pumped by 967 nm diode in the CW mode.

---

* Corresponding author, Tel./fax: +86-551-65591555
  E-mail address: qguo@aiofm.ac.cn

As we know, 2.8 μm laser operated by the transition from $^4I_{11/2}$ to $^4I_{13/2}$ in $Er^{3+}$ is a four-lever system, as shown in Fig.1. The lifetime of the laser upper level $^4I_{11/2}$ is shorter than the lower level $^4I_{13/2}$(See Table 1) in Er:GSGG crystal [17]. Therefore, in principle, the 2.8 μm transition from $^4I_{11/2}$ to $^4I_{13/2}$ is self-terminated. V.Lupei et al established a rate equation model[18] for lasers of $Er^{3+}$ around 2.8 μm, and that a mechanism of major up-conversion process[15] ($W_{11}$, as shown in Fig.1) form $^4I_{13/2}$, ($^4I_{13/2} \rightarrow {}^4I_{15/2}$) +($^4I_{13/2} \rightarrow {}^4I_{9/2}$) under the condition of highly doping $Er^{3+}$ concentration(>30at%) was proposed to overcome the self-terminating effect. The recycling transition(as shown in Fig.1, ~~~›means phonon decay) from $^4I_{9/2} \rightarrow {}^4I_{11/2}$ can increase the population in $^4I_{11/2}$, which is responsible for the formation of inverse population for 2.7-3 μm radiation translation.

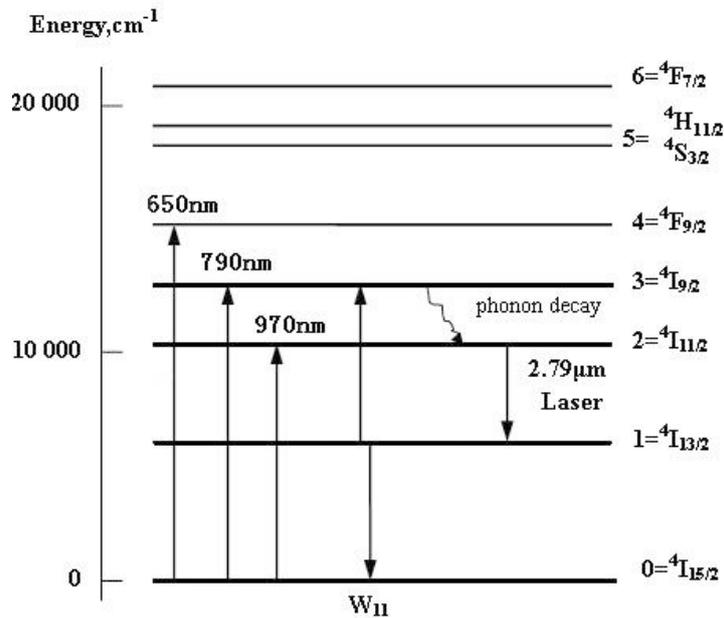

**Fig.1** Energy-Level scheme of $Er^{3+}$ doped crystal

## 2. EXPERIMENTAL STEUP

The experimental configuration is shown in Fig.2. The pumping light with a fiber-coupled output was focused on the 2 mm×2 mm×4 mm, 2 mm×2 mm×5 mm and 3 mm×3 mm×4 mm Er:GSGG (35at% $Er^{3+}$) crystal by coupler lens, respectively. The beam waist diameter in the crystal is approximated to be 0.2 mm. The Er:GSGG crystal was enclosed by a copper heat sink with cooling water passage. Moreover, an indium film was placed between the Er:GSGG crystal and the copper heat sink so that they can be closely contacted. We designed a cavity with lengths of 12 mm, 15 mm and 18 mm, consisting a dielectric input mirror (K9 substrate) highly transmitting at 967 nm and highly reflecting at 2.79 μm and an output mirror (CaF$_2$ substrate) with a reflectivity of 99.5% at 2.79 μm.

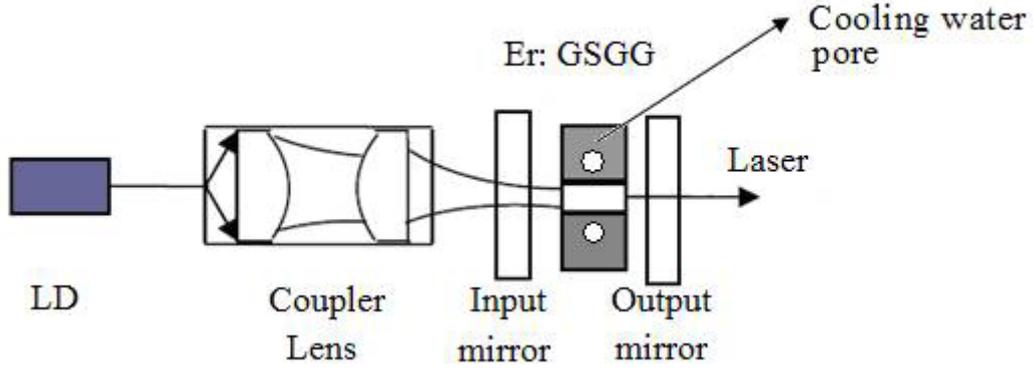

**Fig.2** Schematic experiment setup for the diode end-pumped Er:GSGG laser

## 3 RESULT AND DISCUSSIONS

### 3.1 Numerical simulation

According to the rate equation model from Ref.18, we obtain photon flux density in the laser cavity is:

$$\phi = [\frac{N_0}{\rho}W_p(1-p^2) - \frac{1}{\alpha\sigma N_2} - \frac{W_{22}\rho}{\alpha^2\sigma^2} - N_1(\frac{\beta}{\rho\alpha\tau_2} - \frac{2\beta W_{22}}{\alpha^2\sigma}) - \frac{1-p^2}{\rho\tau_1}N_1]/c \quad (1)$$

$$N_1 = \frac{-1+\sqrt{1+4W_{11}W_p N_0 \tau_1^2}}{2W_{11}\tau_1} \quad (2)$$

$$\rho = [\rho_0 - ln(R_2)]/2l_{eff} \quad (3)$$

$$l_{eff} = l_{cav} + (n-1)l \quad (4)$$

Where $\sigma$ is the emission cross section, $\phi$ is the photon flux density, $N_0, N_1$ and $N_2$ are the populations of level $^4I_{15/2}$, $^4I_{13/2}$ and $^4I_{11/2}$, $\alpha$ and $\beta$ are Boltzmann population coefficients of the level $^4I_{13/2}$ and $^4I_{11/2}$, $W_p$ is the pumping intensity of $^4I_{15/2} \rightarrow ^4I_{11/2}$ (when the crystal is directly pumped by 967 nm diode), $W_{11}$ and $W_{22}$ are energy transfer rates of two-ion up-conversion process of $(^4I_{13/2} \rightarrow {}^4I_{15/2}) + (^4I_{13/2} \rightarrow {}^4I_{9/2})$ and $(^4I_{11/2} \rightarrow {}^4I_{15/2}) + (^4I_{11/2} \rightarrow {}^4S_{3/2})$, $\tau_1$ and $\tau_2$ are life time of level ${}^4I_{13/2}$ and ${}^4I_{11/2}$, $\rho$ is the total losses in resonator, it can be written as equation (3), $R_2$ is the reflective rate of output mirror for 2.79μm laser, $l_{eff}$ is the equivalent length of the cavity and $l$ is the length of the crystal, n is the index of refraction of the Er:GSGG crystal to output laser. All these parameter are shown in Table.1 (The parameters from other References are taken from the data of Er:YSGG[19,20] which are quite similar to those of Er:GSGG).

**Table 1** Part parameters used in the numerical simulation

| $\sigma$ (cm$^2$) | $N_0$ (count/cm$^3$) | $W_1$ (cm$^3$/s) | $W_{22}$ (cm$^3$/s) | $n$ |
|---|---|---|---|---|
| $1.34 \times 10^{-19}$[19] | $4.23 \times 10^{21}$ | $2.3 \times 10^{-16}$[20] | $7.5 \times 10^{-16}$[20] | 1.91 |
| $a$ | $\beta$ | $\tau_1$ (ms) | $\tau_2$ (ms) | |
| 0.294[20] | 0.177[20] | 1.3 | 7.6 | |

Suppose the light in the Er:GSGG crystal is a Gaussian beam in the across area, so the relation between the input power($P_{in}$) and the pumping intensity($W_p$) is:

$$W_p = \frac{2kP_{in}\eta_Q \lambda_p T}{N_0 hc (1-e^{-2}) w(z)^2} \exp(\frac{2r^2}{w(z)^2} - kz) \tag{5}$$

where T (86.8%) is the total transmission of the input mirror and the front surface of the crystal to pumping light, the quantum efficiency $\eta_Q$ (the radio of the number of emitted quanta to the amount of the absorb pumping quanta) is 0.75[20], $k = 8.7\ cm^{-1}$ is the absorption coefficient of the crystal to pumping light, $w(z)$ is the radius of pumping beam at z (suppose the pumping light is focused on the center of the crystal where z=0) while $w_0$ is the beam waist radius, however, $w(z)$ can not be assumed as Gaussian form because the pumping beam has both large waist radius and divergence angle, to be simple, we suppose it change geometrically, therefore it can be written as equation (6) after we simulated the optical path by ZMAX software.

$$w(z) = w_0 + 0.125|z - l/2| \tag{6}$$

The total photon number is

$$\varphi_s = \iiint_V \varphi dV \tag{7}$$

The output power is

$$P_{out} = \frac{c}{2l_{eff}} \varphi_s h v_l \ln(\frac{1}{1-R_2}) \tag{8}$$

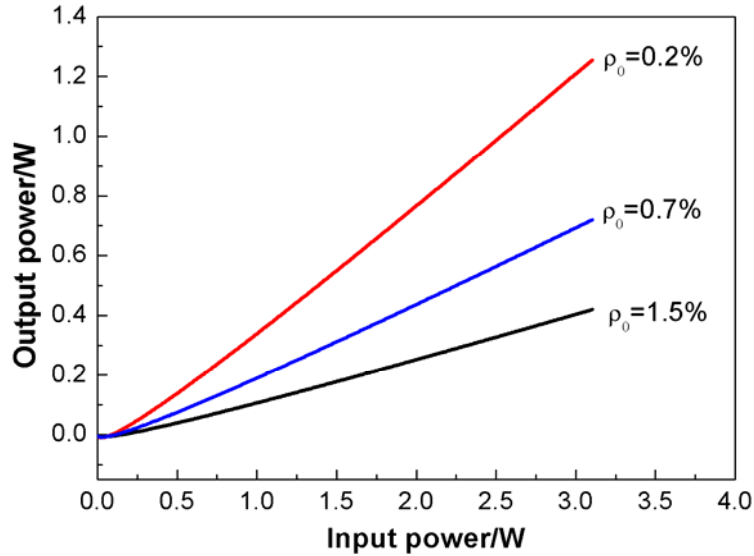

**Fig.3** Output power versus input power according to result of numerical simulation with Er:GSGG crystal of $2mm \times 2mm \times 5mm$ for different values of lost constant $\rho_0$

In the numerical simulation, the length of the crystal is 5 mm, $R_2$ is 99.5%, $l_{cav}$ is 12 mm and w0 is 0.1 mm. The Fresnel number can be written as $N = \frac{a^2}{l\lambda}$, where $a$ is the half length of the crystal which is effect diaphragm here. The value of Fresnel number is 29.87 when the TEM$_{00}$ mode has the lowest lose but others' are much higher [21]. Therefore, to be simple, only the TEM$_{00}$ mode which can almost completely match the resonant cavity is discussed here. The lost constant $\rho_0$ (including absorbing lose, scattered lose, diffractive lose et al in the cavity is relevant to the structure of the resonator. In general, its value is between 0.2%-5% in the diode pumped CW operation. Finally, we obtain the result of numerical simulation with different values of $\rho_0$ (see Fig.3). The highest slope efficiency of 43% with a threshold of 87 mW can be reached by 35at% Er:GSGG crystal if $\rho_0$ can be reduced to 0.2% according to this theoretical model.

## 3.2 Experimental results

The absorption and fluorescence spectra of Er:GSGG are shown in Fig.4. Three absorption bands centered at 650, 790 and 970 nm in the wavelength range of 600-1000 nm are observed, which correspond to the transitions from the ground state $^4I_{15/2}$ to exciting state $^4F_{9/2}$, $^4I_{9/2}$ and $^4I_{11/2}$, respectively. Therefore, 650 nm, 790 nm and 970 nm can be taken as suitable pumping source, as shown in Fig.1. Actually, the 970 nm would have higher conversion efficiency due to it can directly pump the energy to the upper level of 2.79 μm $Er^{3+}$ laser [13]. It can be noted from Fig.4b that many fluorescence peaks (2.64, 2.70, 2.79, 2.82, 2.85, 2.91μm) are observed within 2.6-3.0 μm, resulting from the transitions of stark sub-levels from $^4I_{11/2}$ to $^4I_{13/2}$.

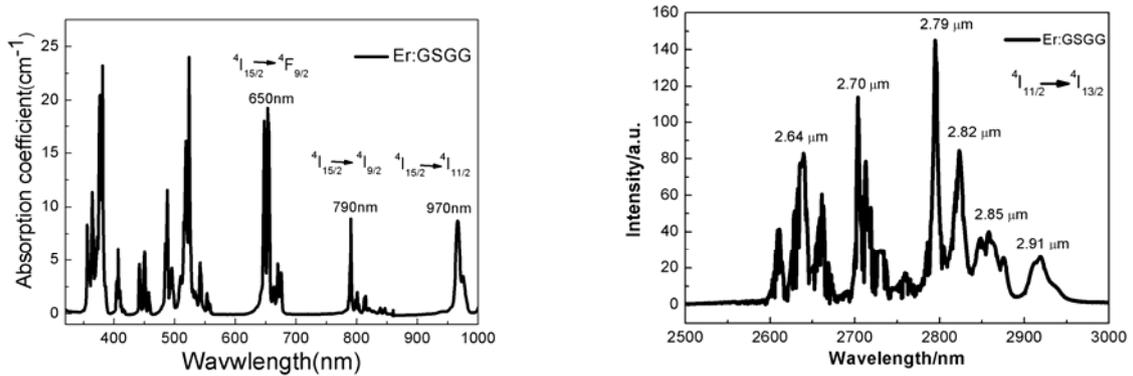

**Fig.4** Spectra of Er:GSGG crystal (a) absorption spectrum.; (b) emission spectrum excited by 967 nm diode

The laser output at 2.79 μm in 35at% Er:GSGG versus the incident 967 nm diode pumping power is illustrated in Fig.5. It can be noted form Fig.5 that a shorter cavity prompts higher output power and slope efficiency. There is a slight oscillation of the output power following the linear fitting line, which goes stronger as the input power increases. The reason may be that the change of the thermal focal length periodically impacts the output power. After we deducted part of pumping power reflected by the input mirror and the front surface of the crystal (about 13.2%), the laser output at 2.79 μm in 35at% Er:GSGG versus the incident 967 nm diode pumping power is illustrated in Fig.5. A maximum output power of 440 mW and a threshold of 139 mW are obtained. The linear fitting results show a slope efficiency of 13.2% and an optical-to-optical efficiency of 13%., which can match with the theoretical values well when $\rho_0$ is 1.5% (the threshold is 135 mW, the slope efficiency is 13.2% and the optical-to-optical efficiency is 12.6%). Using the knife-edge method, it is obtained that the diameter of the beam is 0.9 mm at 18 cm after the output mirror, with the $M^2$ of 2.22 and the divergence angle of 7.2 mrad when the output power is 120 mw.

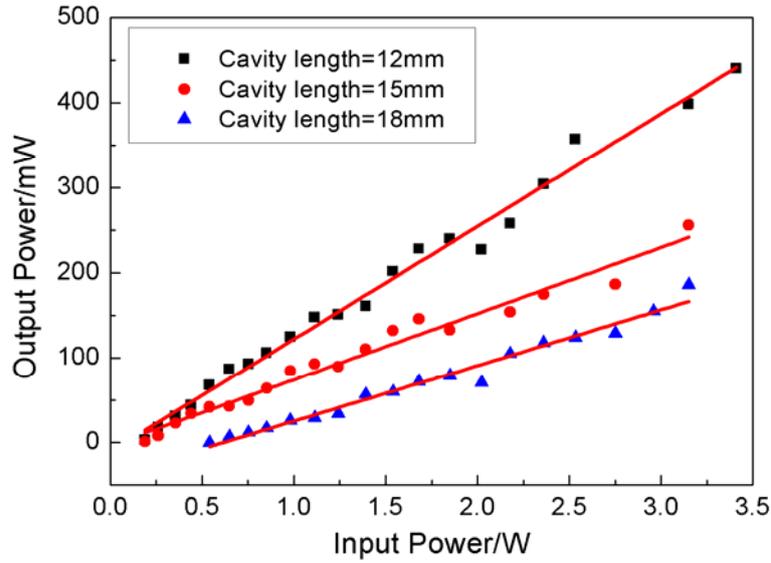

**Fig.5** Laser output at 2.79 μm in 35at% 2 mm×2 mm×5 mm Er:GSGG versus the incident 967 nm diode pumping power under different lengths of the cavity.

To our knowledge, this is the first report of a diode-pumped 2.79 μm Er:GSGG laser with CW mode. Moreover, it has a much higher maximum output power than that of Ti:sapphire pumped 2.79 μm Er:GSGG laser (about 130 mW [15]). However, the slope efficiency is still much lower than that of CW diode pumped 2.79 μm Er:YSGG laser (26% in Ref.4). In our experiment the loss constant $\rho_0$ is still relatively high. Some methods such as a shorter resonant cavity and an improved cooling setup can reduce the loss in the cavity. A more complex and optimized resonant cavity was reported in Ref.13 and Ref.4: first, a dichotic coating with highly transmitting at 967 nm and highly reflecting at 2.79 μm was coated on front surface of crystal as input mirror, which can shorten the length of the cavity to almost the same as the length of the Er:GSGG crystal. Also, a plane-concave cavity was chosen in their experiment. Next, we will design the above cavity structure and optimize the output mirror reflectivity, which is expected to improve the output power and efficiency of diode end-pumped 2.8 μm Er:GSGG laser.

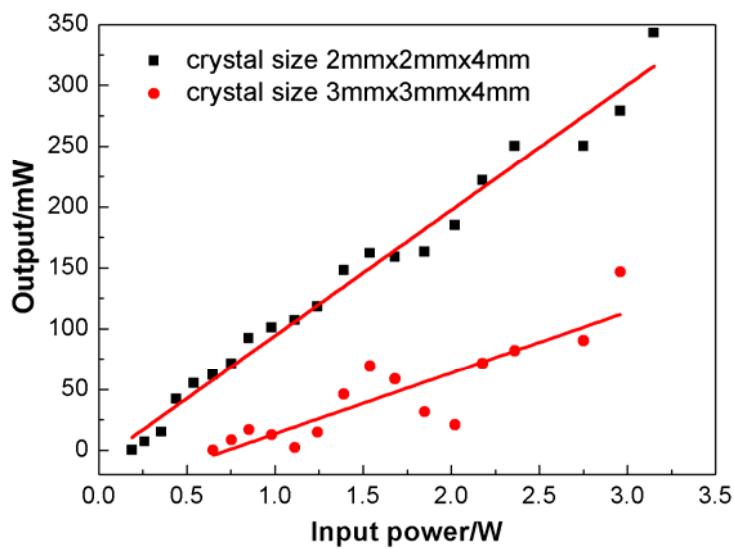

**Fig.6** Laser output at 2.79 μm in 35at% Er:GSGG versus the incident 967 nm diode pumping power with sizes of $2\,mm \times 2\,mm \times 4\,mm$ and $3\,mm \times 3\,mm \times 4\,mm$ when the length of cavity is 12 mm.

To study further about the characteristic of Er:GSGG crystal, we also did experiment with Er:GSGG crystal of $2mm \times 2mm \times 4mm$ and $3\,mm \times 3\,mm \times 4\,mm$ (see Fig. 6). Under the same conditions, a maximum output power of 343 mW and a slope efficiency of 10.3% are obtained by $2\,mm \times 2\,mm \times 4\,mm$ crystal, showing a lower value than that of the $2mm \times 2\,mm \times 5\,mm$ crystal, suggesting the latter has a longer pumping zone so the pumping light can be absorbed more efficiently. Additionally, experiment with Er:GSGG crystal of $3\,mm \times 3\,mm \times 4\,mm$ shows a much worse result that the maximal output power of 146 mW and the slope efficiency of 3.91%. Moreover, the output power has also a much stronger oscillation. In general, the reason may be attributed to a more serious thermal effect because the cross section of $3\,mm \times 3\,mm \times 4\,mm$ crystal is over twice as $2\,mm \times 2\,mm \times 4\,mm$ crystal but the pumping zone is almost the same, leading to a worse cooling effect. As a result, a better cooling in crystal should be an effective way to optimize the laser output as the thermal effect can dramatically influence it.

The specific spectral composition of Er:GSGG laser is illustrated in Fig.7. The laser wavelength ranges from 2.791 μm to 2.798 μm and spectrum width is 3.6 nm. The central wavelength is located at 2.794 μm, suggesting only the strongest fluorescence peak is strengthened and laser oscillation is realized, while others fluorescence peaks are suppressed, as shown in Fig.4.b.

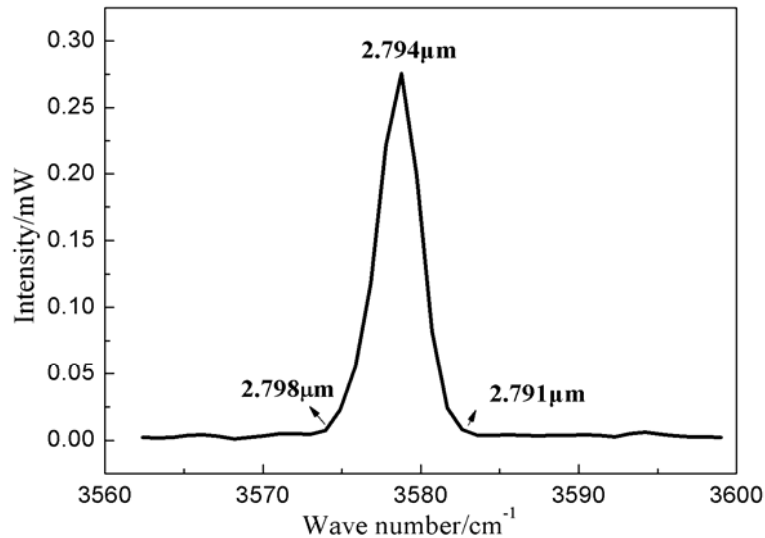

**Fig.7** Spectrum of 2.794 μm Er:GSGG laser pumped by 967 nm diode

## 5. CONCLUSION

We reported a 2.794 μm Er:GSGG laser pumped by 967 nm diode with the maximum output power of 440 mW and spectrum width of 3.6 nm, a slope efficiency of 13.2% and an optical-to-optical efficiency of 13%. The results suggest that short cavity lengths and efficient cooling for crystal are advantageous to increase laser output power. In addition, a optimized output mirror reflectivity, special cavity structure and an advanced cooling system are expected to improve the performance of diode end-pumped Er:GSGG laser.

**ACKNOWLEDGMENTS**

This work was financially supported by National Natural Science Foundation of China under Grant No: 91122021,51272254,61205173 and 90922003).

REFERENCE

1. M. Tempus, W. Luthy, H. P. Weber, V.G. Ostroum, and I.A. Shcherbakov., IEEEJ. Quantum Electron. **30**, 2608 (1994).
2. D.M. Rines, G.A. Rines and P.F. Moulton. Advanced Solid-state Lasers. **24**, 184 (1995).
3. F. Ganikhanov, T. Caughey, K.L.Vodopyanov, J. Opt. Soc. Am. B. **18**, 818 (2001).
4. K.L. Vodopyanov, F. Ganikhanov, J.P. Maffetone, I.Zwieback, W. Ruderman, Optics Letter. **25**, 841 (2000).

5. P.F. MOUTON, IEEE Journal of Quantum Electron. **24**, 960 (1988).

6. T. Jensen, G. Huber, and K. Petermann, Advanced Solid-State Lasers. **1**, 306 (1996).

7. R. S. F. Chang, S. Sengupta, and N. Djeu. Advanced Solid-State Lasers. 162 (1990).

8. T. J. Wang, Q.Y. He, J.Y. Gao, Y. Jiang, Z.H. Kang, H. Sun,L.S. Yu, X.F. Yuan and J. Wu, Laser Physics. **16**,1605 (2006)

9. P. Maak, L. Jakab, P. Richter, H.J. Eichler, and B. liu, Applied Optics. **39,** 3053 (2000).

10. Y.H. Park, D.W. Lee, H.J. Kong, and Y.S. Kim, J. Opt. Soc. Am. B. **25**, 2123, (2008).

11. J. Breguet, A. F. Umyshov, W.A. R. Luthy, I. AShcherbakov, and H.P. Werber , IEEE J.Quantum Electon. **27**, 274 (1991).

12. Bradley, J. Dinerman and P.F. Mouton, Optics Letters. **19**, 1143 (1994).

13. J.S.Liu, J.J.Liu,Y,Tang, Laser Physics. **18**, 1124 (2008).

14. T.J. Wang, Q.Y. He, J.Y. Gao, Z.H. Kang, Y. Jiang, and H. Sun, Laser Phys. Lett. **3**, 349 (2006).

15. R.C. Stoneman, L. Esterowitz, Optics letters. **17**, 816 (1992).

16. D.L. Sun, J.Q. Luo, Q.L. Zhang, J.Z. Xiao, W.P. Liu, S.F. Wang, H.H Jiang, S.T. Yin, Journal of Crystal Growth. **318**, 669 (2011).

17. D.L Sun, J.Q. Luo**,** J.Z. Xiao, Q.L. Zhang, J.K. Chen, W.P. Liu, H.X. Kang, S.T. Yin, Chinese Physics Letter. **29**, 054269 (2012)

18. V. Lupei, S. Georgescu, V. Florea, IEEE J.Quantum Electron. **29**, 426 (1993).

19. E.Arbabzadah, S. Chard, H. Amrania, C.Phillips, and M. Damzen, Optics Express. **19**, 25860 (2011).

20. M. A. Noginov, V.A. Smirnov, I.A. Shcherbakov, Optical and Quantum Electronics. **22**, 61 (1990).

21. Walter Koechner, Solid-state Laser Engineering[M]. Berlin: Springer, 1999.